\documentclass[twovoluments,showpacs,preprintnumbers]{revtex4}
\usepackage{amsmath}
\usepackage{txfonts}
\usepackage[mathscr]{eucal}

\input{tcilatex}
\begin{document}

\title{Hawking radiation from a BTZ black hole viewed as Landauer transport}
\author{ \ Shi-Wei Zhou \  \ Xiao-Xiong Zeng \  \ Wen-Biao Liu (corresponding
author)}
\email{wbliu@bnu.edu.cn}
\affiliation{Department of Physics, Institute of Theoretical Physics, Beijing Normal
University, Beijing, 100875, China}

\begin{abstract}
Viewing Hawking radiation as a 1D single quantum channel Landauer transport
process, Nation et al calculated the energy flux and entropy flux from a
Schwarzschild black hole without chemical potential. To generalize the
method to the case with chemical potential, a rotating charged and
non-charged BTZ black hole is investigated. Energy flux and entropy flux
obtained are consistent with that from anomaly theory. The maximum energy
flux and entropy flux are independent on the statistics of bosons or
fermions.

\end{abstract}

\maketitle

\section{Introduction}

Hawking radiation, which is the most striking quantum effect arising from
the quantum field in a curved spacetime background, attracts many efforts to
investigate it. \ There are several kinds of derivations to obtain it.
Hawking's original derivation is calculating the Bogoliubov coefficients
between the "in" and "out" states in a black hole background \cite{1,2,3}.
The subsequent Damour-Ruffini method is calculating the particles' emitting
rate by analytically extending the outgoing wave from outside of horizon to
inside \cite{4}. Parikh and Wilczek calculated WKB amplitude by considering
pair creation of particles and antiparticles near the horizon and supposing
that particles tunnel across the classical forbidden path where the barrier
is created by the outgoing particles themselves \cite{6,7,8,9}. Another
approach to calculate Hawking radiation is due to anomaly through
calculating the energy-momentum tensor in the black hole backgrounds \cite%
{10,11,12,13,14}. The anomaly in field theory occurs if the symmetry of the
action or the corresponding conservation law is valid in the classical
theory but will violate in the quantized case.\ Anomalies can include
conformal anomaly (or trace anomaly), anomaly in gauge symmetries, and
gravitation anomaly. The pioneer work of Christensen and Fulling \cite{10}
told us that the strength of Hawking radiation flux is determined by the
trace of energy-momentum tensor. Thinking of the two dimensional massless
field, either Hawking effect or conformal anomaly can be deduced from the
other. In Robinson and Wilczek's work \cite{11}, Hawking radiation can be
understood as compensating flux to cancel gravitation anomaly at the horizon.

 In 80's, Zurek has viewed  Hawking radiation as a 3D black body radiation obeying Stefan-Boltzman law \cite{33}.
 However, some recent  works is indicating that a 4D black hole metric can be reduced to (1+1) dimensional spacetime
 and defined flat Rindler spacetime by virtue of the conformal symmetry near the horizon ,
 thus Hawking radiation is inherently  a (1+1) dimensional process .
Recently, a new 1D single quantum channel transport model was used to
explain Hawking radiation \cite{15}. This model was first proposed to
measure the electronic conductance of electrical transport in mesoscopic
physics and was subsequently extended to thermal transport. It has been
proved that the thermal conduction of 1D ballistic transport  based on
fractional statistics and the Landauer formulation is independent on the
statistics nature and is governed by the universal quantum $\kappa ^{univ}=%
\frac{\pi ^{2}}{3}\frac{k_{B}^{2}T}{h}$ in the degenerate regime \cite%
{16,17,18,19}. In Ref.\cite{15}, Hawking radiation energy and entropy flow
of a Schwarzschild black hole is viewed as a 1D single quantum transport
process. For a Schwarzschild black hole, the chemical potential $\mu _{BH}=0$%
, the flux of bosons such as photons and gravitons is equal to the result of
1D quantum transport in the degenerate limit. For fermions such as neutrinos
and electrons, in order to get the same maximum flux, the bi-direction
current of particles and antiparticles must be taken into account.
Meanwhile, the universal upper bound $\dot{S}_{1D}^{2}\leq \big(\frac{\pi
k_{B}^{2}}{3\hbar }\big)\dot{E}_{1D}$ holds all the time. The result
obtained from 1D quantum channel model is consistent with that from
conformal symmetry arising near the horizon of (1+1)-dimensional
Schwarzschild black hole. In Refs.\cite{20,21}, the method has been extended
to some more complicated black holes with nonzero chemical potential and it
is found that charge flux and gauge flux can also be viewed as a 1D Landauer
transport process.

(2+1)-dimensional BTZ black hole as a solution of standard Einstein field
equation $G_{ab}+\Lambda g_{ab}=\kappa T_{ab}$ with negative cosmological
constant $\Lambda =-l^{-2}$ \cite{22,23,24,25}, has attracted great
attention because it provides a simplified model for exploring black hole
thermodynamics and quantum gravity. Hawking radiation from a BTZ black hole
has been investigated in Refs.\cite{26,27,28}. We will generalize Nation's
method to the case of nonzero chemical potential of the rotating charged BTZ
black hole and a special case with $Q=0$ in this paper.

\section{The metric of a BTZ black hole}

The line element of a charged rotating BTZ black hole can be written as \cite%
{22,23}%
\begin{equation}
ds^{2}=-N^{2}(r)dt^{2}+N^{-2}(r)dr^{2}+r^{2}\big[N^{\varphi }(r)dt+d\varphi %
\big]^{2},  \label{2}
\end{equation}%
where the squared lapse\ $N^{2}(r)$\ and the angular shift \ $N^{\varphi
}(r) $\ are given as%
\begin{equation}
N^{2}(r)=-M+\frac{r^{2}}{l^{2}}+\frac{J^{2}}{4r^{2}}-\frac{\pi }{2}Q^{2}\ln
r\equiv f(r),  \label{1a}
\end{equation}
\begin{equation}
N^{\varphi }(r)=-\frac{J}{2r^{2}},  \label{1b}
\end{equation}%
with $-\infty <t<\infty $, $0<r<\infty $, and $0\leq \varphi \leq 2\pi $.
The black hole is characterized by the parameters ADM mass $M$, angular
momentum $J$, and electronic charge $Q$ carried by the black hole, which
determine the asymptotic behavior of the solution. The metric is stationary
and axially symmetric with Killing vectors $\partial _{t}$ and $\partial
_{\varphi }$.

Horizons of the charged rotating BTZ black hole are roots of the lapse
function. We only care the case while two distinct real roots exist, then
the following inequality should be satisfied%
\begin{equation}
M>\frac{\pi Q^{2}+\sqrt{\pi ^{2}Q^{4}+\frac{16J^{2}}{l^{2}}}}{8}+\frac{2J^{2}%
}{l^{2}\bigg(\pi Q^{2}+\sqrt{\pi ^{2}Q^{4}+\frac{16J^{2}}{l^{2}}}\bigg)}.
\label{a}
\end{equation}%
Hawking temperature is%
\begin{equation}
T_{BH}=\frac{1}{4\pi }f^{^{\prime }}(r)\big|_{r_{+}}=\frac{1}{2\pi r_{+}}%
\bigg(\frac{r_{+}^{2}}{l^{2}}-\frac{J^{2}}{4r_{+}^{2}}-\frac{\pi Q^{2}}{4}%
\bigg).  \label{6}
\end{equation}

Especially, $Q=0$ is corresponding to a non-charged rotating BTZ black hole,
and then the outer and inner event horizons $r_{\pm }$ are given as%
\begin{equation}
r_{\pm }^{2}=\frac{l^{2}}{2}\bigg(M\pm \sqrt{M^{2}-\frac{J^{2}}{l^{2}}}\bigg)%
.  \label{3}
\end{equation}%
In terms of the inner and outer horizons, the black hole mass and angular
momentum are%
\begin{equation}
M=\frac{r_{+}^{2}}{l^{2}}+\frac{J^{2}}{4r_{+}^{2}}=\frac{r_{+}^{2}+r_{-}^{2}%
}{l^{2}},\qquad J=\frac{2r_{+}r_{-}}{l},  \label{4}
\end{equation}%
with the corresponding angular velocity%
\begin{equation}
\Omega =\frac{J}{2r^{2}}.  \label{5}
\end{equation}%
Hawking temperature is%
\begin{equation}
T_{BH}=\frac{1}{4\pi }f^{^{\prime }}(r)\big|_{r_{+}}=\frac{1}{2\pi r_{+}}%
\bigg(\frac{r_{+}^{2}}{l^{2}}-\frac{J^{2}}{4r_{+}^{2}}\bigg).  \label{5a}
\end{equation}

For the region near the horizon of a rotating BTZ black hole, the quantum
field can be effectively described by an infinite collection of
(1+1)-dimensional fields. The Kaluza-Klein reduction of the
(2+1)-dimensional BTZ black hole yields%
\begin{equation}
ds^{2}=-f(r)dt^{2}+f^{-1}(r)dr^{2},  \label{7}
\end{equation}%
with a $U(1)$ gauge field%
\begin{equation}
A_{t}=-\frac{J}{2r^{2}}-\frac{1}{2}Q^{2}\ln r,\qquad A_{t}=-\frac{J}{2r^{2}},
\label{8a}
\end{equation}%
and they are corresponding to the charged and uncharged circumstances
respectively.

\section{Hawking radiation flux calculated using anomalies}

Hawking radiation from a rotating BTZ black hole can be obtained from gauge
and gravitational anomalies \cite{30}. For a reduced two-dimensional metric
Eq.(\ref{7}), the gravitational anomaly of the chiral scalar field is%
\begin{equation}
\nabla _{\mu }T_{\nu }^{\mu }=\frac{1}{96\pi \sqrt{-g}}\epsilon ^{\beta
\delta }\partial _{\delta }\partial _{\alpha }\Gamma _{\nu \beta }^{\alpha },
\label{9}
\end{equation}%
which can also be rewritten as%
\begin{equation}
\nabla _{\mu }T_{\nu }^{\mu }\equiv A_{\nu }=\frac{1}{\sqrt{-g}}\partial
_{\mu }N_{\nu }^{\mu },  \label{10}
\end{equation}%
where $N_{\nu }^{\mu }$ is defined as%
\begin{equation}
N_{\nu }^{\mu }=\frac{1}{96\pi }\epsilon ^{\beta \mu }\partial _{\alpha
}\Gamma _{\nu \beta }^{\alpha },  \label{11}
\end{equation}%
and the epsilon tensor reads%
\begin{equation}
\epsilon ^{\mu \nu }=\left(
\begin{array}{cc}
0 & 1 \\
-1 & 0%
\end{array}%
\right) .  \label{12}
\end{equation}%
After calculating, non-zero components of $N_{\nu }^{\mu }$ are%
\begin{equation}
N_{t}^{r}=\frac{1}{192\pi }(f^{^{\prime }2}+f^{^{\prime \prime }}f),
\label{13}
\end{equation}%
\begin{equation}
N_{r}^{t}=-\frac{1}{192\pi f^{2}}(f^{^{\prime }2}-f^{^{\prime \prime }}f).
\label{14}
\end{equation}

Now considering the gauge anomaly, neglecting classically irrelevant ingoing
modes near the horizon, the effective two-dimensional theory becomes chiral
near the horizon and the gauge symmetry becomes anomalous.

Outside far from the horizon, the current is conserved with%
\begin{equation}
\partial _{r}J_{(o)}^{r}=0,  \label{15}
\end{equation}%
while in the region near the horizon, since there is only outgoing field,
the current satisfies the anomalous equation as%
\begin{equation}
\partial _{r}J_{(H)}^{r}=\frac{m^{2}}{4\pi }\partial _{r}A_{t},  \label{16}
\end{equation}%
where $m$ is the $U(1)$ charge.

By integrating Eq.(\ref{15}) and Eq.(\ref{16}), one can obtain the flux in
each region as%
\begin{equation}
J_{(o)}^{r}=C_{0},\qquad J_{(H)}^{r}=C_{H}+\frac{m^{2}}{4\pi }%
[A_{t}(r)-A_{t}(r_{+})],  \label{15a}
\end{equation}%
where $C_{0}$ and $C_{H}$ are integration constants. Under gauge
transformations, variation of effective action is%
\begin{equation}
-\delta W=\int d^{2}x\sqrt{-g_{2}}\lambda \nabla _{\mu }J^{\mu },
\label{17a}
\end{equation}%
where $\lambda $ is a gauge parameter and%
\begin{equation}
J^{\mu }=J_{(o)}^{\mu }\theta _{+}(r)+J_{(H)}^{\mu }H(r),  \label{17b}
\end{equation}%
where $\theta _{+}(r)=\theta (r-r_{+}-\epsilon )$, $H(r)=1-\theta _{+}(r)$.

Using the anomaly equation, we get%
\begin{equation}
-\delta W=\int d^{2}x\lambda \bigg[\delta (r-r_{+}-\epsilon )\bigg(%
J_{(o)}^{r}-J_{(H)}^{r}+\frac{m^{2}}{4\pi }A_{t}\bigg)+\partial _{r}\bigg(%
\frac{m^{2}}{4\pi }A_{t}H\bigg)\bigg].  \label{17c}
\end{equation}%
The total effective action should be gauge invariant and the last term
should be canceled by quantum effects of the classically irrelevant ingoing
modes . The quantum effect to cancel this term is induced by the ingoing
modes near the horizon. The coefficient of the delta-function should vanish,
so we have%
\begin{equation}
C_{0}=C_{H}-\frac{m^{2}}{4\pi }A_{t}(r_{+}).  \label{17d}
\end{equation}%
Since the covariant current is written as ${\tilde{J^{r}}}=J^{r}+\frac{m^{2}%
}{4\pi }A_{t}H$, the condition ${\tilde{J^{r}}}=0$ determines the value of
the charge flux to be%
\begin{equation}
C_{0}=-\frac{m^{2}}{4\pi }A_{t}(r_{+}).  \label{17e}
\end{equation}%
Similarly the total flux of the energy-momentum tensor can be obtained as%
\begin{equation}
a_{0}=\frac{m^{2}}{4\pi }A_{t}^{2}(r_{+})+N_{t}^{r}(r_{+}).  \label{17e1}
\end{equation}%
As for the charged rotating BTZ black hole with Eq.(\ref{1a}), the gauge
potential $A_{t}$ is given by Eq.(\ref{8a}) and $N_{t}^{r}$ is given by Eq.(%
\ref{13}). Thus the gauge flux $C_{0}$ and total energy flux $a_{0}$ are
written as%
\begin{equation}
C_{0}=\frac{m^{2}}{4\pi }\bigg(\frac{J}{2r_{+}^{2}}+\frac{1}{2}Q^{2}\ln r_{+}%
\bigg),\qquad a_{0}=\frac{m^{2}}{4\pi }\bigg(\frac{J}{2r_{+}^{2}}+\frac{1}{2}%
Q^{2}\ln r_{+}\bigg)^{2}+\frac{\pi }{12}T_{BH}^{2}.  \label{17e2}
\end{equation}%
Letting $Q=0$ will give the case of uncharged BTZ black hole. We can find
that gauge and gravitational anomalies in a BTZ black hole can be canceled
by the total flux of Hawking radiation at Hawking temperature.

\section{Landauer transport model for Hawking radiation from a black hole}

Now we will consider a single channel connecting two particle/heat
reservoirs with (quasi-)particles obeying fractional statistics (generalized
Bose and Fermi statistics) proposed by Haldane \cite{31,32}. The two
reservoirs are characterized by the temperatures $T_{L}$ and $T_{R}$ with
chemical potential $\mu _{L}$ and $\mu _{R}$, respectively. The subscripts $%
L $ and $R$ denote the left and right reservoirs respectively while we
assume that $T_{L}>T_{R}$ and the transport through 1D connection is
adiabatic and ballistic (no scattering).

For particles obeying fractional statistics, the distribution function is
given by%
\begin{equation}
f_{g}(x)=\frac{1}{\omega (x,g)+g},  \label{18a}
\end{equation}%
with $\omega (x,g)$ given by the implicit function equation%
\begin{equation}
\omega ^{g}(x,g)[1+\omega (x,g)]^{1-g}=e^{x},  \label{18b}
\end{equation}%
where $x\equiv \beta (E-\mu )$, $\beta \equiv \frac{1}{k_{B}T}$, $g$ is the
statistics parameter satisfying $g\geq 0$, $g=0$ and $g=1$ describe bosons
and fermions respectively.

The net energy flux $\dot{E}$ based on Landauer theory is $\dot{E}=\dot{E}%
_{L}-\dot{E}_{R}$ with%
\begin{equation}
\dot{E}_{L(R)}=\frac{1}{h}\int_{E_{L(R)}^{(0)}}^{\infty }Ef_{g}^{L(R)}dE,
\label{18c}
\end{equation}%
where we have taken the canceling of the group velocity and density of state
into account, and the particle transmission probability is supposed as 1.

Changing the integration variable from $E$ to $x=\beta (E-\mu )$, we have%
\begin{equation}
\dot{E}_{L(R)}=\frac{(k_{B}T_{L(R)})^{2}}{2\pi \hbar }\int_{x_{L(R)}^{0}}^{%
\infty }dx\big(x+\frac{\mu _{L(R)}}{k_{B}T_{L(R)}}\big)f_{g}^{L(R)}(x).
\label{18d}
\end{equation}%
Thinking of the fermion case, when the contribution of antiparticles is
considered, the maximum energy flux of fermions is written as%
\begin{equation}
\dot{E}_{L(R)}=\frac{(k_{B}T_{L(R)})^{2}}{2\pi \hbar }\bigg[\int_{\frac{-\mu
_{L(R)}}{k_{B}T_{L(R)}}}^{\infty }dx\big(x+\frac{\mu _{L(R)}}{k_{B}T_{L(R)}}%
\big)\frac{1}{e^{x}+1}+\int_{\frac{\mu _{L(R)}}{k_{B}T_{L(R)}}}^{\infty }dy%
\big(y+\frac{\mu _{L(R)}}{k_{B}T_{L(R)}}\big)\frac{1}{e^{y}+1}\bigg].
\label{18e}
\end{equation}%
It can be rewritten as%
\begin{align}
\dot{E}_{L(R)}& =\frac{(k_{B}T_{L(R)})^{2}}{2\pi \hbar }\bigg[%
\int_{0}^{\infty }dx\big(x+\frac{\mu _{L(R)}}{k_{B}T_{L(R)}}\big)\frac{1}{%
e^{x}+1}+\int_{0}^{\frac{\mu _{L(R)}}{k_{B}T_{L(R)}}}dx\big(-x+\frac{\mu
_{L(R)}}{k_{B}T_{L(R)}}\big)\frac{1}{e^{-x}+1}  \notag \\
& +\int_{0}^{\infty }dy\big(y+\frac{\mu _{L(R)}}{k_{B}T_{L(R)}}\big)\frac{1}{%
e^{y}+1}-\int_{0}^{\frac{\mu _{L(R)}}{k_{B}T_{L(R)}}}dy\big(y+\frac{\mu
_{L(R)}}{k_{B}T_{L(R)}}\big)\frac{1}{e^{y}+1}\bigg]  \notag \\
& =\frac{(k_{B}T_{L(R)})^{2}}{2\pi \hbar }\bigg[\int_{0}^{\infty }dx\frac{x}{%
e^{x}+1}+\int_{0}^{\infty }dy\frac{y}{e^{y}+1}+2\frac{\mu _{L(R)}}{%
k_{B}T_{L(R)}}\bigg(\int_{0}^{\infty }dx\frac{1}{e^{x}+1}-\int_{0}^{\frac{%
\mu _{L(R)}}{k_{B}T_{L(R)}}}dx\frac{1}{e^{x}+1}\bigg)+\frac{\mu _{L(R)}^{2}}{%
2(k_{B}T_{L(R)})^{2}}\bigg],  \label{18g}
\end{align}%
so we have%
\begin{equation}
\dot{E}=\dot{E}_{L}-\dot{E}_{R}=\frac{\pi k_{B}^{2}}{12\hbar }%
(T_{L}^{2}-T_{R}^{2})+\frac{1}{4\pi \hbar }(\mu _{L}^{2}-\mu _{R}^{2}),
\label{18h}
\end{equation}%
where we have considered that the upper limit of integral $\frac{\mu _{L(R)}%
}{k_{B}T_{L(R)}}$ approaches to infinity.

For the charge flux, we have%
\begin{equation}
\dot{I}=\frac{k_{B}T_{L(R)}e}{2\pi \hbar }\int_{\frac{-\mu _{L(R)}}{%
k_{B}T_{L(R)}}}^{\infty }dx\frac{1}{e^{x}+1}.  \label{18i}
\end{equation}%
Considering the contribution of antiparticles, we get%
\begin{equation}
\dot{I}=\frac{k_{B}T_{L(R)}e}{2\pi \hbar }\int_{\frac{-\mu _{L(R)}}{%
k_{B}T_{L(R)}}}^{\infty }dx\frac{1}{e^{x}+1}+\frac{k_{B}T_{L(R)}e}{2\pi
\hbar }\int_{\frac{\mu _{L(R)}}{k_{B}T_{L(R)}}}^{\infty }dy\frac{1}{e^{y}+1}.
\label{19}
\end{equation}%
Similarly, in the degenerate limit, the lower integration bound $\frac{\mu
_{L(R)}}{k_{B}T_{L(R)}}$ approaches to infinity, so we obtain%
\begin{equation}
\dot{I}=\frac{e}{2\pi \hbar }(\mu _{L}-\mu _{R}).  \label{20}
\end{equation}

As for the case of bosons, in the limit of degeneration, similar calculation
as the fermion case can give the same conclusion as Eq.(\ref{18h}).

According to Ref.\cite{17}, the net entropy flux $\dot{S}_{1D}$ is $\dot{S}%
_{1D}=\dot{S}_{L}-\dot{S}_{R}$ with%
\begin{equation}
\dot{S}_{L(R)}=-\frac{k_{B}^{2}T_{L(R)}}{2\pi \hbar }\int_{x_{L(R)}^{0}}^{%
\infty }dx\bigg[f_{g}\ln f_{g}+(1-gf_{g})\ln (1-gf_{g})-\big[1+(1-g)f_{g}%
\big]\ln \big[1+(1-g)f_{g}\big]\bigg].  \label{18c1}
\end{equation}%
Changing integration variable $x=\beta (E-\mu )$ to $\omega $, the entropy
flux can be simplified to%
\begin{equation}
\dot{S}_{L(R)}=\frac{k_{B}^{2}T_{L(R)}}{2\pi \hbar }\int_{\omega _{g}\big(%
\frac{-\mu _{L(R)}}{k_{B}T_{L(R)}}\big)}^{\infty }d\omega \bigg[\frac{\ln
(1+\omega )}{\omega }-\frac{\ln \omega }{1+\omega }\bigg].  \label{18c2}
\end{equation}%
In the degenerate limit, the lower integration bound approaches to zero,
therefore the statistics-dependence vanishes. The maximum entropy flux can
be obtained%
\begin{equation}
\dot{S}_{1D}=\frac{\pi k_{B}^{2}}{6\hbar }(T_{L}-T_{R}).  \label{18c3}
\end{equation}

Till now, we have obtained the energy flux and charge flux for a 1D Landauer
transport process. Taking Hawking radiation from a black hole as Landauer
transport, where one reservoir is black hole with Hawking temperature $%
T_{L}=T_{BH}$ and black hole's electronic chemical potential $\mu _{L}=\mu
_{BH}$, the other reservoir is vacuum with $T_{R}=\mu _{R}=0,$ we can give
the total energy flux and charge flux as%
\begin{equation}
\dot{E}=\dot{E}_{L}-\dot{E}_{R}=\frac{\pi k_{B}^{2}}{12\hbar }T_{BH}^{2}+%
\frac{1}{4\pi \hbar }\mu _{BH}^{2},\qquad \dot{I}=\frac{e}{2\pi \hbar }\mu
_{BH}.  \label{18h1}
\end{equation}%
where $\mu _{BH}=mA_{t}$. This is consistent with Eq.(\ref{17e2}). The
entropy flux is%
\begin{equation}
\dot{S}_{1D}=\frac{\pi k_{B}^{2}}{6\hbar }T_{BH}.  \label{18Z3}
\end{equation}

In fact, when the 1D quantum transport system can be viewed as a
near-equilibrium one, the electric flux ($\dot{I}$) and energy flux ($\dot{E}
$) yield \cite{18}%
\begin{equation}
\delta \dot{I}=\frac{\partial \dot{I}}{\partial \mu }\bigg|_{T}\delta \mu +%
\frac{\partial \dot{I}}{\partial T}\bigg|_{\mu }\delta T,\qquad \delta \dot{E%
}=\frac{\partial \dot{E}}{\partial \mu }\bigg|_{T}\delta \mu +\frac{\partial
\dot{E}}{\partial T}\bigg|_{\mu }\delta T.  \label{23}
\end{equation}%
with $\delta T=T_{L}-T_{R}$, $\delta \mu =\mu _{L}-\mu _{R}$.

For a system of fractional statistics under Eq.(\ref{18a}), taking the limit
$\delta T\rightarrow 0$ and $\delta \mu \rightarrow 0$, the linear transport
coefficients for arbitrary $g>0$ can be given as \cite{18}%
\begin{equation}
L_{11}=\frac{\partial \dot{I}}{\partial \mu }\bigg|_{T}=M\frac{e}{2\pi \hbar
}\int_{0}^{\infty }\frac{d\omega }{(\omega +g)^{2}}=M\frac{e}{2\pi \hbar }%
\frac{1}{g},  \label{25}
\end{equation}

\begin{equation}
L_{12}=\frac{\partial \dot{I}}{\partial T}\bigg|_{\mu }=M\frac{e}{2\pi \hbar
}k_{B}\int_{0}^{\infty }d\omega \frac{x(\omega ,g)}{(\omega +g)^{2}}=0,
\label{26}
\end{equation}

\begin{equation}
L_{21}=\frac{\partial \dot{E}}{\partial \mu }\bigg|_{T}=M\frac{1}{2\pi \hbar
\beta }\int_{0}^{\infty }d\omega \frac{x(\omega ,g)+\mu \beta }{(\omega
+g)^{2}}=M\frac{\mu }{2\pi \hbar }\frac{1}{g},  \label{27}
\end{equation}
\begin{equation}
L_{22}=\frac{\partial \dot{E}}{\partial T}\bigg|_{\mu }=M\frac{k_{B}^{2}T}{%
2\pi \hbar }\int_{0}^{\infty }d\omega \frac{x^{2}(\omega ,g)+\mu \beta
x(\omega ,g)}{(\omega +g)^{2}}=M\frac{k_{B}^{2}T}{2\pi \hbar }\frac{\pi ^{2}%
}{3},  \label{28}
\end{equation}%
where $M$ is an integer related to the occupied modes number, we will
neglect it by letting it to be 1.

So the energy flux and charge flux are%
\begin{equation}
\delta \dot{E}=\frac{\mu }{2\pi \hbar }\frac{1}{g}\delta \mu +\frac{%
k_{B}^{2}T}{2\pi \hbar }\frac{\pi ^{2}}{3}\delta T,\qquad \delta \dot{I}=%
\frac{e}{2\pi \hbar }\frac{1}{g}\delta \mu .  \label{29a}
\end{equation}%
It also means that%
\begin{equation}
\dot{E}_{L(R)}=\frac{\mu _{_{L(R)}}^{2}}{4\pi \hbar g}+\frac{\pi k_{B}^{2}}{%
12\hbar }T_{L(R)}^{2},\qquad \dot{I}=\frac{e}{2\pi \hbar }\frac{1}{g}\mu .
\label{29b}
\end{equation}

\section{Conclusion and discussion}

Following Nation's work, we have generalized it to the case with chemical
potential. Viewing Hawking radiation as a 1D Landauer transport, we have
calculated the energy flux and charge flux from a rotating charged and
uncharged BTZ black hole. The total energy flux obtained from Landauer
transport model, which is consistent with that from anomaly theory, contains
not only thermal flux but also the contribution of flux caused by chemical
potential. Based on the fractional statistics, the energy flux and entropy
flux are independent on statistical behavior in the degenerate regime. From
the above formulism Eqs.(\ref{29a}) and (\ref{29b}), we can easily find that
the total energy flux of a 1D quantum transport system in the degenerate
limit can be divided into two parts: the first term is due to the difference
of chemical potential of the two reservoirs; the second term is purely
thermal and entirely determined by the temperature. In addition, setting $%
\delta \mu =0$, the total energy flux will eliminate to net thermal flux
generated by $\delta T$. The 1D thermal conductance $\kappa ^{univ}=\frac{%
\pi ^{2}}{3}\frac{k_{B}^{2}T}{h}$ is exactly the coefficient $L_{22}$, which
is independent on statistical behavior. As for the electronic flux, $%
L_{12}=0 $ means that it only depends on the chemical potential.

It is noticeable that Hawking radiation is viewed as a phenomenon  near the horizon . 
As for infinity , The Hawking radiation  would contain a gray-body factor , 
which  caused by the effective potential outside the horizon .

\begin{acknowledgments}
This research is supported by the National Natural Science Foundation of
China under Grant Nos. 10773002 and 10875012. It is also supported by the
Scientific Research Foundation of Beijing Normal University under Grant No.
105116.
\end{acknowledgments}

\end{document}